\documentclass[aps,amsmath,amssymb,prb,preprint]{revtex4}

\usepackage{graphicx}
\usepackage{dcolumn}
\usepackage{bm}
\usepackage[dvips]{color} 

\begin{document}

\title{Spin rectification induced by dynamical Hanle effect}

\author{Hiroto Sakimura}
\affiliation{Department of Applied Physics and Physico-Informatics, Keio University, Yokohama 223-8522, Japan}

\author{Takahiko Matsumoto}
\affiliation{Department of Applied Physics and Physico-Informatics, Keio University, Yokohama 223-8522, Japan}

\author{Kazuya Ando\footnote{ando@appi.keio.ac.jp}}
\affiliation{Department of Applied Physics and Physico-Informatics, Keio University, Yokohama 223-8522, Japan}

\begin{abstract}
Dynamic response of spin accumulation to a time-dependent magnetic field has been investigated in a ferromagnetic/nonmagnetic bilayer under ferromagnetic resonance. In this system, magnetization precession driven by a microwave generates direct-current (dc) and alternate-current (ac) spin accumulation in the nonmagnetic layer by the spin pumping. The ac spin accumulation is coupled with the microwave magnetic field through a dynamical Hanle spin precession, giving rise to rectified spin accumulation comparable with the dc spin accumulation directly generated by the spin pumping. 
\end{abstract}

\maketitle

Rectification effects are fundamental in electrical, optical, and magnetic systems. An electrical rectifier is used to convert an alternating current to a direct current. The rectifier essentially strips the high-frequency or alternating part from the incoming current and delivers a low-frequency current, which is based on the trigonometric relation: $\cos\omega _1 t  \cos \omega_2 t=(1/2)[\cos(\omega_1-\omega_2) t+\cos(\omega_1+\omega_2)t]$; when two waves of frequencies $\omega_1$ and $\omega_2$ are combined, the difference-frequency and sum-frequency terms appear. If the frequencies satisfy $\omega_1=\omega_2$, the term $\cos(\omega_1-\omega_2) t$ gives rise to a direct-current (dc) signal. A similar dynamic response to the product of alternate-current (ac) spin accumulation and a time-dependent magnetic field is the origin of a spin rectification effect induced by a dynamical Hanle effect presented in this paper.

The Hanle effect refers to the variation of spin accumulation $\delta {\bf m}$ in response to a magnetic field ${\bf H}$ applied perpendicular to the spin-polarization direction~\cite{Jedema,Dash,Fukuma}; when a nonmagnet is exposed to a magnetic field, spins in the nonmagnet start to precess around ${\bf H}$ as $\partial \delta {\bf m}/\partial t=-\gamma \delta {\bf m}\times {\bf H}$.

\begin{figure}[bt]
\includegraphics[scale=1]{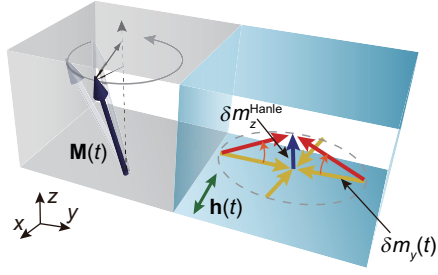}
\caption{A schematic illustration of the spin pumping and dynamical Hanle effect. ${\bf M}(t)$ and ${\bf h}(t)$ are the magnetization and microwave magnetic field, respectively. The spin pumping creates ac spin accumulation $\delta m_y(t)$, which is rectified through the spin precession induced by ${\bf h}(t)$, giving rise to dc spin accumulation $\delta m_z^\text{Hanle}$. 
}
\label{fig1} 
\end{figure}

\begin{figure}[bt]
\includegraphics[scale=1]{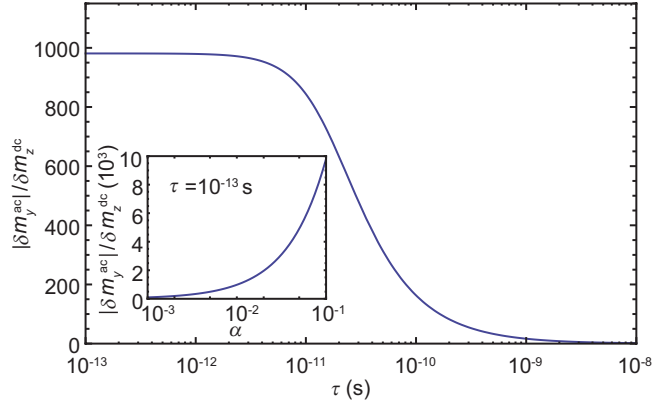}
\caption{The ratio ${{\left| {{\delta m_y}(t)} \right|}}/{{\delta m_z^{{\rm{dc}}}}}$ of the magnitude of the ac spin accumulation $\left| {{\delta m_y}(t)} \right|$ to that of the dc spin accumulation ${{\delta m_z^{{\rm{dc}}}}}$ directly induced by the spin pumping as a function of the spin relaxation time $\tau$ in the nonmagnetic layer for $4\pi M_s=0.745$ T, $h=0.01$ mT, $\alpha=0.01$, and $f= 9.4$ GHz. The inset shows Gilbert damping constant $\alpha$ dependence of ${{\left| {{\delta m_y}(t)} \right|}}/{{\delta m_z^{{\rm{dc}}}}}$. 
}
\label{fig2} 
\end{figure}

In this letter, we show rectification of ac spin accumulation through spin precession induced by a time-dependent magnetic field: a dynamical Hanle effect. The dynamical Hanle effect is discussed in a ferromagnetic/nonmagnetic ($F/N$) junction under ferromagnetic resonance, where uniform magnetization precession is driven by a microwave magnetic field. In this system, the magnetization precession generates dc and ac spin accumulation, or spin currents, in the $N$ layer by the spin pumping.~\cite{PhysRevLett.110.217602} The dc-component of the spin pumping has been intensely studied in recent years.~\cite{Saitoh,Costache,andoAPL,ando:262505,PhysRevLett.104.046601,PhysRevB.83.144402,PhysRevLett.107.046601} In contrast, the ac-component of the spin pumping has been directly observed only recently.~\cite{2013arXiv1307.2961W} Here, we show that the ac-component of the spin pumping not only generates an ac electric voltage through the inverse spin Hall effect but also generates, through the dynamical Hanle spin precession, large dc spin accumulation comparable with the dc spin accumulation directly generated by the dc-component of the spin pumping. This finding offers a way to generate a large dc electric voltage from the ac-component of the spin pumping using the inverse spin Hall effect.

The spin pumping in a $F/N$ junction generates a spin current ${{\bf{j}}_s}(t)$ and spin accumulation $\delta {\bf m}(t)$ in the $N$ layer.~\cite{Tserkovnyak1,Brataas,Mizukami,Tserkovnyak2,Heinrich,Saitoh,Costache,Taniguchi,andoAPL,ando:262505,PhysRevLett.104.046601,PhysRevB.83.144402,PhysRevLett.107.046601,PhysRevLett.110.217602} The spin current density created by the spin pumping is expressed as~\cite{Tserkovnyak1,Brataas}
\begin{equation}
{{\bf{j}}_s}(t) = \frac{\hbar }{{4\pi }}g_\text{eff}^{\uparrow\downarrow}\frac{1}{{M_s^2}}\left( {{\bf{M}}(t) \times \frac{{d{\bf{M}}}(t)}{{dt}}} \right),\label{pump}
\end{equation}
where $g_\text{eff}^{\uparrow\downarrow}$ is the spin pumping conductance and $M_s$ is the saturation value of magnetization ${\bf M}$. Equation~(\ref{pump}) shows that the spin pumping generates two types of spin currents; a dc spin current with the spin polarization direction along the $z$ axis, i.e. the magnetization precession axis, and an ac spin current whose spin polarization direction oscillates in the $x$-$y$ plane under the ferromagnetic resonance (FMR) condition [see Fig.~\ref{fig1}]. In a $F/N$ thin film, where the magnetocrystalline anisotropy can be neglected and a microwave magnetic field ${\bf h}(t)=(h \cos \omega t,0,0)$ is applied along the $x$ axis, the dc and ac spin currents with the spin polarization direction along $z$ and $y$ axis at the resonance condition are obtained from Eq.~(\ref{pump}) with the Landau-Lifshitz-Gilbert equation by ignoring the second-order contribution of the precession amplitude as~\cite{AndoJAPfull}
\begin{equation}
{j_{s}^z} = \frac{{g_\text{eff}^{\uparrow\downarrow}{h^2}{\gamma ^2}\hbar \left( {2\pi {M_s}\gamma  + \sqrt {4{\pi ^2}{M_s}^2{\gamma ^2} + {\omega ^2}} } \right)}}{{16\pi {\alpha ^2}\left( {4{\pi ^2}{M_s}^2{\gamma ^2} + {\omega ^2}} \right)}},
\end{equation}
\begin{equation}
{j_{s}^y}(t) = {j_{s1}^y}\cos \omega t - {j_{s2}^y}\sin \omega t, \label{jsy1}
\end{equation}
where 
\begin{equation}{j_{s1}^y} = \frac{{g_\text{eff}^{\uparrow\downarrow}h\gamma \hbar \left( {2\pi {M_s}\gamma  + \sqrt {4{\pi ^2}{M_s}^2{\gamma ^2} + {\omega ^2}} } \right)}}{{8\pi \alpha \sqrt {4{\pi ^2}{M_s}^2{\gamma ^2} + {\omega ^2}} }},
\end{equation}  
\begin{equation}
{j_{s2}^y} = \frac{{g_\text{eff}^{\uparrow\downarrow}h\gamma \hbar  \omega }}{{8\pi  \sqrt {4{\pi ^2}{M_s}^2{\gamma ^2} + {\omega ^2}} }}.
\end{equation}
Here, $\gamma$ and $\alpha$ are the gyromagnetic ratio and the Gilbert damping constant, respectively. In the $N$ layer, the dynamics of spin accumulation $\delta {\bf m}(x,t)=(\delta m_x(t), \delta m_y(t), \delta m_z(t))$ induced by the spin pumping is obtained from~\cite{AndoNMad}
\begin{equation}
\frac{{\partial {\delta\bf{m}}(x,t)}}{{\partial t}} =  - \gamma  {{\delta\bf{m}}(x,t) \times {{\bf{H}}_{{\rm{eff}}}}}(t)  - \frac{{{\delta\bf{m}}(x,t)}}{\tau } + D{\nabla ^2}{\delta\bf{m}}(x,t) + {{\bf{j}}_s}(t)\delta (x),\label{acc}
\end{equation}
where ${\bf H}_\text{eff}$ is the effective magnetic field, $\tau$ is the spin relaxation time, and $D$ is the diffusion coefficient in the $N$ layer. For simplicity, in the following discussions, we neglect the external magnetic field and consider the system where the $N$ layer is thin enough so that the spin diffusion term can be neglected.

The magnitude of the ac and dc spin accumulation directly generated by the spin pumping, i.e., the spin accumulation in the absence of the spin precession: $  {{\delta\bf{m}} \times {{\bf{H}}_{{\rm{eff}}}}} ={\bf 0}$, depends critically on the spin relaxation time $\tau$ in the $N$ layer. In the absence of the spin precession, the ac spin accumulation ${\delta m_y}(t) $ induced by the spin pumping is obtained from Eqs.~(\ref{jsy1}) and (\ref{acc}) as
\begin{equation}
{\delta m_y}(t) = \frac{\tau }{{1 + {{(\omega \tau )}^2}}}\left[ {({j_{s1}^y} + \omega \tau {j_{s2}^y})\cos \omega t + (\omega \tau {j_{s1}^y} - {j_{s2}^y})\sin \omega t} \right].\label{deltamy}
\end{equation}
Using Eq.~(\ref{deltamy}), we plot the ratio of the ac to dc spin accumulation, ${{\left| {{\delta m_y}(t)} \right|}}/{{\delta m_z^{{\rm{dc}}}}}$, in Fig.~\ref{fig2}, where the dc spin accumulation $\delta m_z^{{\rm{dc}}}$ directly generated by the spin pumping is obtained from Eq.~(\ref{acc}): ${{d{\delta m_z}^\text{dc}}}/{{dt}} =0=  - {\delta m_z}^\text{dc}/{\tau } + {j_{s}^z}$. Here, $4\pi M_s=0.745$ T, $h=0.01$ mT, $\gamma=1.86\times 10^{11}$ T$^{-1}$s$^{-1}$, $\alpha=0.01$, and $f= 9.4$ GHz were used for the calculation, where $\omega=2\pi f$ and $f$ is the microwave frequency. Figure~\ref{fig2} shows that ${{\left| {{\delta m_y}(t)} \right|}}/{{\delta m_z^{{\rm{dc}}}}}$ changes critically around $\tau\simeq 1/\omega=1.7\times 10^{-11}$ s. In a material with $\tau\gg 1/\omega$, where the time scale of the oscillation of the ac spin accumulation $1/\omega$ is much faster than the spin relaxation time $\tau$, the ac spin accumulation induced by the spin pumping is almost averaged out, whereas the dc spin accumulation can be accumulated in the time scale of $\tau$. In contrast, in a material with $\tau\ll 1/\omega$, the spins injected into the $N$ layer relaxes before cancelling out the ac spin accumulation. Here note that the ac spin current $j_s^y$ is much larger than the dc spin current $j_s^z$, since the cone angle of the magnetization precession is typically less than 1$^\circ$, resulting the large ac spin accumulation ${{\left| {{\delta m_y}(t)} \right|}}$ compared with the dc spin accumulation ${{\delta m_z^{{\rm{dc}}}}}$ in this system. In fact, as shown in the inset to Fig.~\ref{fig2}, ${{\left| {{\delta m_y}(t)} \right|}}$ increases with the Gilbert damping constant $\alpha$, showing that the ac spin accumulation is significant at small amplitudes of magnetization precession, since the precession amplitude decreases with increasing $\alpha$.

\begin{figure}[bt]
\includegraphics[scale=1]{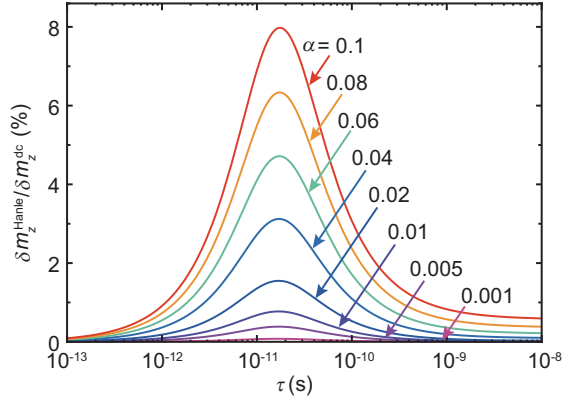}
\caption{The ratio ${{\delta m_z^{{\rm{Hanle}}}}}/{{\delta m_z^{{\rm{dc}}}}}$ of the dc spin accumulation induced by the Hanle effect $\delta m_z^{{\rm{Hanle}}}$ to that induced by the spin pumping $\delta m_z^\text{dc}$ calculated for different Gilbert damping constants $\alpha$ with $4\pi M_s=0.745$ T, $h=0.01$ mT, and $f= 9.4$ GHz. 
}
\label{fig3} 
\end{figure}

\begin{figure}[bt]
\includegraphics[scale=1]{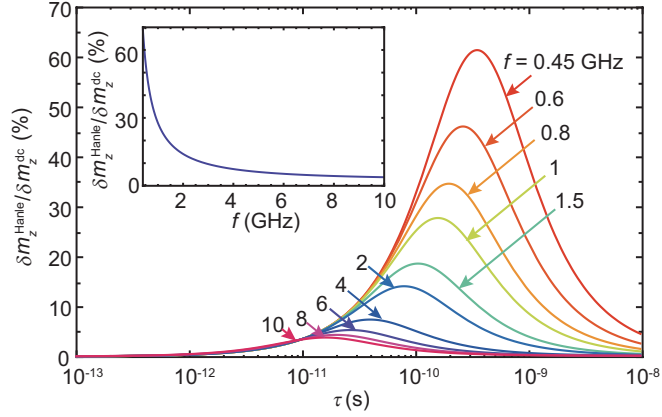}
\caption{${{\delta m_z^{{\rm{Hanle}}}}}/{{\delta m_z^{{\rm{dc}}}}}$ calculated for different microwave frequencies $f$ as a function of the spin relaxation time $\tau$. $4\pi M_s=0.745$ T, $h=0.01$ mT, and $\alpha=0.05$ were used for the calculation. The inset shows $f$ dependence of the maximum value of ${{\delta m_z^{{\rm{Hanle}}}}}/{{\delta m_z^{{\rm{dc}}}}}$, i.e., ${{\delta m_z^{{\rm{Hanle}}}}}/{{\delta m_z^{{\rm{dc}}}}}$ at $\tau=1/\omega$. 
}
\label{fig4} 
\end{figure}

The ac spin accumulation ${\delta m_y}(t)$ induced by the spin pumping creates rectified spin accumulation $\delta m_z^{{\rm{Hanle}}}$ with the spin polarization direction along the $z$ axis through the dynamical Hanle effect; ${\delta m_y}(t)$ is rectified by the applied microwave magnetic field ${\bf h}(t)=(h \cos \omega t,0,0)$ because of the spin-precession term in Eq.~(\ref{acc}): $  - \gamma  {{\delta\bf{m}}(x,t) \times {{\bf{h}}(t)}}$. By neglecting a contribution from $\delta m_z$ on $\delta m_y$, the $z$ component of the spin accumulation due to the precession of $\delta m_y(t)$ induced by ${\bf h}(t)$ is obtained from ${{d{\delta m_z}(t)}}/{{dt}} = \gamma {\delta m_y}(t)h\cos \omega t - {\delta m_z}(t)/{\tau }$. The spin precession term $\gamma \delta m_y (t) h\cos \omega t$ with the term $\delta m_y (t)\propto \cos \omega t$ gives rise to rectified dc spin accumulation $\delta m_z^{{\rm{Hanle}}} = ({\omega }/{{2\pi }})\int_0^{2\pi /\omega } {{\delta m_z}(t)} dt$: 
\begin{equation}
\delta m_z^{{\rm{Hanle}}} = \frac{{h\gamma {\tau ^2}({j_{s1}^y} + \omega \tau {j_{s2}^y})}}{{2\left[ {1 + {{(\omega \tau )}^2}} \right]}}.
\end{equation}
In Fig.~\ref{fig3}, we show the ratio of the dc spin accumulation $\delta m_z^{{\rm{Hanle}}}$ induced by the dynamical Hanle effect to the dc spin accumulation $\delta m_z^\text{dc}=\tau j_s^z $ directly induced by the dc spin pumping: 
\begin{equation}
\frac{{\delta m_z^{{\rm{Hanle}}}}}{{\delta m_z^{{\rm{dc}}}}} = \frac{{\alpha \tau \left( {4{\pi ^2}{M_s}^2{\gamma ^2} + {\omega ^2}} \right)\left[ {2\pi {M_s}\gamma  + \alpha \tau {\omega ^2} + \sqrt {4{\pi ^2}{M_s}^2{\gamma ^2} + {\omega ^2}} } \right]}}{{\left[ {1 + {{(\omega \tau )}^2}} \right]\left[ {{\omega ^2} + 2\pi {M_s}\gamma \left( {2\pi {M_s}\gamma  + \sqrt {4{\pi ^2}{M_s}^2{\gamma ^2} + {\omega ^2}} } \right)} \right]}}\label{rat}
\end{equation}
for $4\pi M_s=0.745$ T and $f= 9.4$ GHz. Figure \ref{fig3} shows that ${{\delta m_z^{{\rm{Hanle}}}}}/{{\delta m_z^{{\rm{dc}}}}}$ is maximized around $\tau\simeq 1/\omega=1.7\times 10^{-11}$ s. Here note that $\tau\simeq 1/\omega$ is the spin relaxation time where the ratio of the ac to dc spin accumulation, ${{\left| {{\delta m_y}(t)} \right|}}/{{\delta m_z^{{\rm{dc}}}}}$, changes drastically [see Fig.~\ref{fig2}], suggesting that it is important to consider separately the two situations: $\tau \ll 1/\omega$ and $\tau \gg 1/\omega$. In a material with $\tau \ll 1/\omega$, spins injected into the $N$ layer relax before the precession of $\delta m_y(t)$ due to ${\bf h}(t)$, resulting the suppression of $\delta m_z^\text{Hanle}$ as Fig.~\ref{fig3}. In contrast, in a material with $\tau \gg 1/\omega$, $\delta m_z^\text{Hanle}$ is suppressed by the suppression of ${{\left| {{\delta m_y}(t)} \right|}}/{{\delta m_z^{{\rm{dc}}}}}$ [see Fig.~\ref{fig2}], resulting the peak structure of ${{\delta m_z^{{\rm{Hanle}}}}}/{{\delta m_z^{{\rm{dc}}}}}$ with respect to $\tau$. Therefore, the dynamical Hanle effect is maximized in a material with the spin relaxation time of $\tau\simeq 1/\omega$, showing that the rectification can be controlled by tuning the microwave frequency $f=\omega/(2\pi)$.

The rectification effect is also sensitive to the Gilbert damping constant $\alpha$ of the $F$ layer as shown in Fig.~\ref{fig3}; the dynamical Hanle spin precession is efficient in a system with large $\alpha$. This is due to the fact that the ac spin accumulation $\delta m_y(t)$ induced by the spin pumping becomes significant compared with the dc spin accumulation $\delta m_z^{{\rm{dc}}} $ with increasing $\alpha$ as shown in the inset to Fig.~\ref{fig2}. At first sight, this result seems to imply that the spin rectification depends also on the amplitude of the microwave magnetic field $h$, since the cone angle of the magnetization precession decreases with decreasing $h$, or ${{\left| {{\delta m_y}(t)} \right|}}/{{\delta m_z^{{\rm{dc}}}}}$ increases with decreasing $h$. However, this compensates the decrease of the spin-precession term; by decreasing $h$, the spin-precession term $  {{\delta\bf{m}} \times {{\bf{H}}_{{\rm{eff}}}}} $ also decreases, resulting that the rectification of the spin accumulation is independent of $h$ as described in Eq.~(\ref{rat}).

The magnitude of the rectified spin accumulation ${\delta m_z^{{\rm{Hanle}}}}$ can be comparable with that of the dc spin accumulation ${{\delta m_z^{{\rm{dc}}}}}$ directly induced by the dc spin pumping at low microwave frequencies. Figure~\ref{fig4} shows ${{\delta m_z^{{\rm{Hanle}}}}}/{{\delta m_z^{{\rm{dc}}}}}$ at different microwave frequencies $f$ calculated using Eq.~(\ref{rat}) with $4\pi M_s=0.745$ T and $\alpha=0.05$. Figure~\ref{fig4} shows that the spin relaxation time $\tau_\text{max}$ at which ${\delta m_z^{{\rm{Hanle}}}}$ is maximized increases with decreasing $f$ because of $\tau_\text{max}\simeq 1/\omega$. Notably, the maximum value of ${{\delta m_z^{{\rm{Hanle}}}}}/{{\delta m_z^{{\rm{dc}}}}}$ increases with decreasing $f$ [see also the inset to Fig.~\ref{fig4}], and ${{\delta m_z^{{\rm{Hanle}}}}}$ exceeds 50\% of ${{\delta m_z^{{\rm{dc}}}}}$ at $f=0.45$ GHz. Since the rectification depends on the magnetic damping $\alpha$ as shown in Fig.~\ref{fig3}, ${\delta m_z^{{\rm{Hanle}}}}$ can be further enhanced by selecting a spin injector ferromagnet with a large magnetic damping $\alpha$, providing a route for realizing ${{\delta m_z^{{\rm{Hanle}}}}}/{{\delta m_z^{{\rm{dc}}}}}> 100\%$.

In summary, we have shown that ac spin accumulation generated by the spin pumping is rectified through spin precession induced by a microwave magnetic field: the dynamical Hanle effect. The magnitude of the rectified spin accumulation can be comparable with that of the dc spin accumulation directly induced by the spin pumping. The rectification is sensitive to the Gilbert damping constant and microwave frequency, providing a route for generating giant dc spin accumulation through the dynamical Hanle effect.

This work was supported by the Cabinet Office, Government of Japan through its Funding Program for Next Generation World-Leading Researchers, the Asahi Glass Foundation, the Noguchi Institute, the Murata Science Foundation, and the Mitsubishi Foundation.

\end{document}